\documentclass{PoS}
\usepackage{bm,amsmath,amssymb}
\usepackage{cite}

\newcommand{\rmd}{{\rm d}}
\newcommand{\rme}{{\rm e}}

\newcommand{\bk}{{\bm{k}}}

\newcommand{\bp}{{\bm{p}}}
\newcommand{\bx}{{\bm{x}}}
\newcommand{\by}{{\bm{y}}}
\newcommand{\bu}{\bm{u}}

\newcommand{\bz}{{\bm{z}}}

\newcommand{\asbar}{\bar{\alpha}_s}

\title{Resummation of Large Logarithms in the Rapidity Evolution of Color Dipoles}

\ShortTitle{Large Logs Resummation in BK Equation}

\author{E. Iancu $^a$, \speaker{J.D. Madrigal} $^{a}$, A.H. Mueller $^{b}$, G. Soyez $^a$, and D.N. Triantafyllopoulos $^c$ \\ 
\llap{$^a$}Institut de Physique Théorique\\ 
CEA Saclay, UMR 3681, F-91191 Gif-sur-Yvette, France.\\ 
\llap{$^b$}Department of Physics, Columbia University\\ 
New York, NY 10027, USA.\\ 
\llap{$^c$}European Centre for Theoretical Studies in Nuclear Physics and Related Areas (ECT*) and \\Fondazione Bruno Kessler\\ 
Strada delle Tabarelle 286, I-38123 Villazzano (TN), Italy.\\ 
E-mail: \email{edmond.iancu@cea.fr}, \email{jose-daniel.madrigal-martinez@cea.fr}, 
\email{amh@phys.columbia.edu}, \email{gregory.soyez@cea.fr}, \email{trianta@ectstar.eu}.}

\abstract{Perturbative corrections beyond leading-log accuracy to BFKL and BK equations, describing the rapidity evolution of QCD scattering amplitudes at high energy, exhibit strong convergence problems due to radiative corrections enhanced by large single and double transverse logs. We identify explicitly the physical origin of double transverse logs and resum them directly in coordinate space as appropriate for BK equation, in terms of an improved local-in-rapidity evolution kernel. Numerical results show the crucial role of double-logarithmic resummation for BK evolution, which is stabilized and slowed down by roughly a factor of two. }

\FullConference{XXIII International Workshop on Deep-Inelastic Scattering,\\
		27 April - May 1 2015\\
		Dallas, Texas}

\begin{document}

\section{Introduction}

Along the last two decades, the next-to-leading (NLO) corrections to the rapidity evolution of QCD scattering amplitudes in the asymptotic high-energy regime, given by the Balitsky-Fadin-Kuraev-Lipatov (BFKL) \cite{BFKL} equation and its nonlinear extensions, the Balitsky-Kovchegov (BK) \cite{BK} and Jalilian-Marian-Iancu-McLerran-Weigert-Leonidov-Kovner (JIMWLK) \cite{JIMWLK} equations, have become available \cite{NLOBFKL,NLOBK1,NLOBK2,NLOJIMWLK}. Going beyond leading logarithmic accuracy is known to be mandatory in order to have a chance to accurately describe the available precision data on DIS at small Bjorken-$x$ or forward $pA$ collisions (for a review, see e.g. \cite{kl,cgc}).\\

However, it is known that the size of the NLO corrections to the BFKL kernel is very large, threatening to deprive the formalism of its predictiveness \cite{ross}. The source for this poor convergence was identified as due to large transverse collinear logarithms appearing at each order in the perturbative expansion starting at NLO, and successful all-orders resummation procedures in Mellin space were devised \cite{salam} by effectively adding the higher-order terms involving poles in the characteristic function required by DGLAP evolution in collinear and anticollinear regions\footnote{In the work \cite{sabio}, this procedure was approximately reformulated as an improvement at the level of the BFKL kernel in momentum space.}. The convergence problem is not expected to disappear when including nonlinear effects (multiple scattering and saturation) as encoded in the BK equation \cite{dion}, a fact that was confirmed explicitly in a recent numerical study \cite{lappi,ours}.\\

Implementing the collinear improvement at the level of the non-linear BK equation is the major step carried out in the work \cite{ours}. For this it is needed to identify the origin of the large corrections directly in coordinate space. Previous works in this direction \cite{motyka,beuf} had already recognized the interplay between the \emph{kinematical constraint} \cite{kinematical} and double collinear logs. In \cite{ours}, we derived this relationship by a thorough analysis of Feynman diagrams in the double logarithmic approximation (Sec. \ref{2}). Although time ordering naturally provides a non-local evolution equation  \eqref{eq:nonlocal}, we were able to show that rather nontrivially the double-log resummation can be implemented at the level of the (energy-independent) kernel, rendering its applicability more feasible than that of former approaches (Sec. \ref{3}). Having reformulated the resummation in terms of a modified BK kernel, we were able to show numerically its big impact on both stabilizing and slowing down the evolution (Sec. \ref{4}). This entails important consequences for phenomenology: in a recent analysis we show how resummed BK evolution provides an excellent fit to the combined HERA data for the DIS structure functions at small $x_{\rm Bjorken}$ with a small number of fitting parameters having values in the physically expected ballpark \cite{fit}.

\section{Double Logarithmic Approximation in High-Energy QCD Evolution}\label{2}
The probability for gluon emission with transverse momentum $k_\perp$ and a fraction $x$ of the parent parton longitudinal momentum is enhanced for soft and collinear emissions: ${\rm d}{\cal P}\sim\bar{\alpha}_s\frac{{\rm d}x}{x}\frac{{\rm d}k_\perp^2}{k_\perp^2}\equiv\bar{\alpha}_s{\rm d}Y{\rm d}\rho$, where $\bar{\alpha}_s=\frac{\alpha_s N_c}{\pi}$, $Y=\ln(1/x)$ is the rapidity variable and $\rho=\ln(k_\perp^2/Q_0^2)$, with $Q_0$ the characteristic transverse scale of the target. In the kinematic region where $\bar{\alpha}_sY\rho\gtrsim 1$ while $\bar{\alpha}_s Y\ll 1$ and $\bar{\alpha}_s\rho\ll 1$, the resummation of terms of the form $(\bar{\alpha}_sY\rho)^n$ to all orders is expected to provide a reasonable approximation. This \emph{naive} double-logarithmic approximation (DLA) appears in the BFKL equation, describing the rapidity evolution of the dipole S-matrix $S_{\bx\by}=\frac{1}{N_c}{\rm tr}[V_\bx^\dagger V_\by]=1-T_{\bx\by}$
\begin{equation}\label{bfkleq}
 \partial_YT_{\bx\by}(Y)=\frac{\asbar}{2\pi}\int\rmd^2\bz\,{\cal M}_{\bx\by\bz}[T_{\bx\bz}(Y)+T_{\bz\by}(Y)-T_{\bx\by}(Y)];\quad{\cal M}_{\bx\by\bz}=\frac{(\bx-\by)^2}{(\bx-\bz)^2(\bz-\by)^2},
\end{equation}
by taking the limit where the daughter dipoles are much larger than the original one, $|\bx-\bz|\simeq|\bz-\by|\gg r\equiv|\bx-\by|$. In this limit, averaging over angles and impact parameter $T_{\bx\by}(Y)\to T(Y,r^2)$ and making the rescaling $T(Y,r^2)\equiv r^2Q_0^2{\cal A}(Y,r^2)$, Eq. \eqref{bfkleq} reduces to
\begin{equation}\label{naive}
 {\cal A}(Y,r^2)={\cal A}(0,r^2)+\asbar\int_0^Y\rmd Y_1\int_{r^2}^{1/Q_0^2}\frac{\rmd z^2}{z^2}{\cal A}(Y_1,z^2),
\end{equation}
which resums by iteration the terms enhanced by factors ($\asbar Y\rho)^n$. A crucial point to notice, however, is that these are not the only terms enhanced with the maximum number of large logs per power of the coupling. In fact, we will see that kinematical constraints also generate terms like $\asbar\rho^2,\asbar^2Y\rho^2,\asbar^2\rho^4,\cdots$ which can be seen as large corrections to the impact factor or formally higher-order corrections to the BFKL kernel, but will become important and need to be resummed when $Y>\rho\gtrsim 1/\sqrt{\asbar}$.
\begin{figure}[t]
\centering
\includegraphics[scale=0.65]{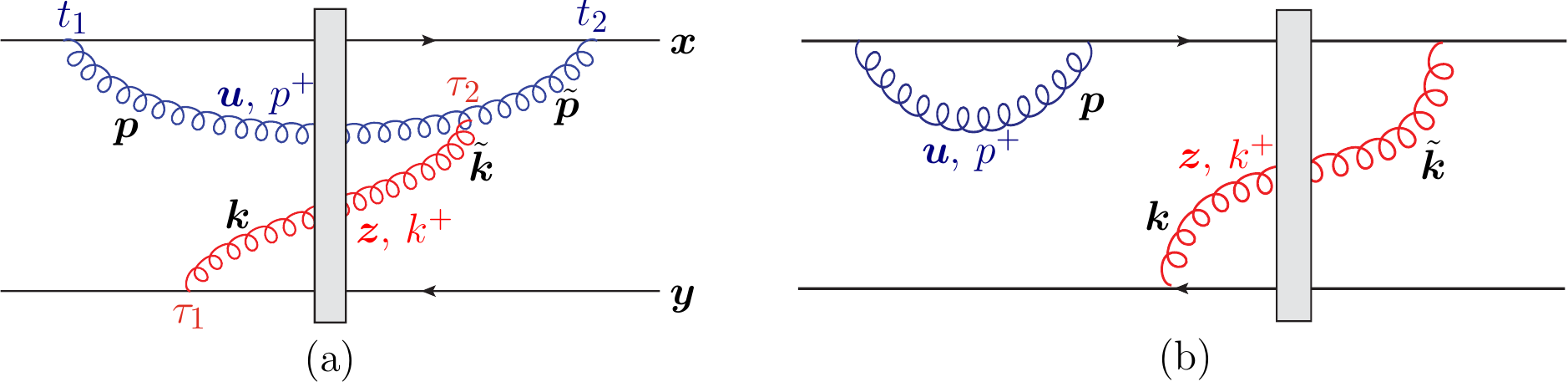}
\caption{Sample diagrams with two gluon ordered in longitudinal momentum ($p^+\gg k^+$) and also in lifetime ($\tau_p>\tau_k$): real-real graph (a); virtual-real graph (b).}
\label{fig1}
\end{figure}
To see how these higher powers of the transverse logs are generated, we consider a diagram like (a) in Fig. \ref{fig1}. It is proportional to
\begin{equation}
 \int_{q_0^+}^{q^+}\frac{\rmd k^+}{k^+}\int^{q^+}_{k^+}\frac{\rmd p^+}{p^+}\frac{p^+}{p^++k^+\frac{\bp^2}{\bk^2}}\frac{p^+}{p^++k^+\frac{(\tilde{\bp}-\tilde{\bk})^2}{\tilde{\bk}^2}},\quad Y=\ln(q^+/q_0^+).
\end{equation}
The factor $\frac{p^+}{p^++k^+\frac{\bp^2}{\bk^2}}$ is equal to $\frac{\tau_p}{\tau_p+\tau_k}$, where $\tau_p\equiv 1/p^-=2p^+/\bp^2$ is the associated gluon lifetime. To double log accuracy this can be replaced simply by $\Theta(\tau_p-\tau_k)$, and moreover this time ordering can be enforced directly in coordinate space \cite{ours}. Then, the DLA approximation to the sum of real-real graphs for NLO color dipole evolution reads
\begin{equation}
 \begin{aligned}
  \left(\frac{\asbar}{2\pi}\right)^2\int_{q_0^+}^{q^+}\frac{\rmd k^+}{k^+}\int_{k^+}^{q^+}\frac{\rmd p^+}{p^+}\int_{\bu\bz}{\cal M}_{\bx\by\bu}[{\cal M}_{\bu\by\bz}S_{\bx\bu}S_{\bu\bz}S_{\bz\by}+{\cal M}_{\bx\bu\bz}S_{\bx\bz}S_{\bz\bu}S_{\bu\by}]\Theta(p^+\bar{u}^2-k^+\bar{z}^2),
 \end{aligned}
\end{equation}
with $\bar{u}={\rm max}(|\bu-\bx|,|\bu-\by|);\, \bar{z}={\rm max}(|\bz-\bx|,|\bz-\by|)$. As we saw in arriving to \eqref{naive}, the region
\begin{equation}
 |\bz-\bx|\simeq|\bz-\by|\simeq|\bz-\bu|\gg|\bu-\bx|\simeq\bu-\by|\gg r=|\bx-\by|
\end{equation}
generates the double logs. Indeed, within this approximation ${\cal M}_{\bu\by\bz}{\cal M}_{\bx\by\bu}\simeq\frac{r^2}{\bar{u}^2\bar{z}^4}$ and $1-S_{\bx\bu}S_{\bu\bz}S_{\bz\by}\simeq T_{\bu\bz}+T_{\bz\by}\simeq2T(\bar{z}^2)$, which gives rise to the double logs through
\begin{equation}
 \int_{r^2}^{\bar{z}^2}\frac{\rmd\bar{u}^2}{\bar{u}^2}\int_{k^+\frac{\bar{z}^2}{\bar{u}^2}}^{q^+}\frac{\rmd p^+}{p^+}=\int_{r^2}^{\bar{z}^2}\frac{\rmd\bar{u}^2}{\bar{u}^2}\left(\ln\frac{q^+}{k^+}-\ln\frac{\bar{z}^2}{\bar{u}^2}\right)=Y\rho-\frac{\rho^2}{2}; \quad Y=\ln\frac{q^+}{k^+},\,\rho=\ln\frac{\bar{z}^2}{r^2}.
\end{equation}
\begin{figure}[t]
\centering
\includegraphics[scale=0.7]{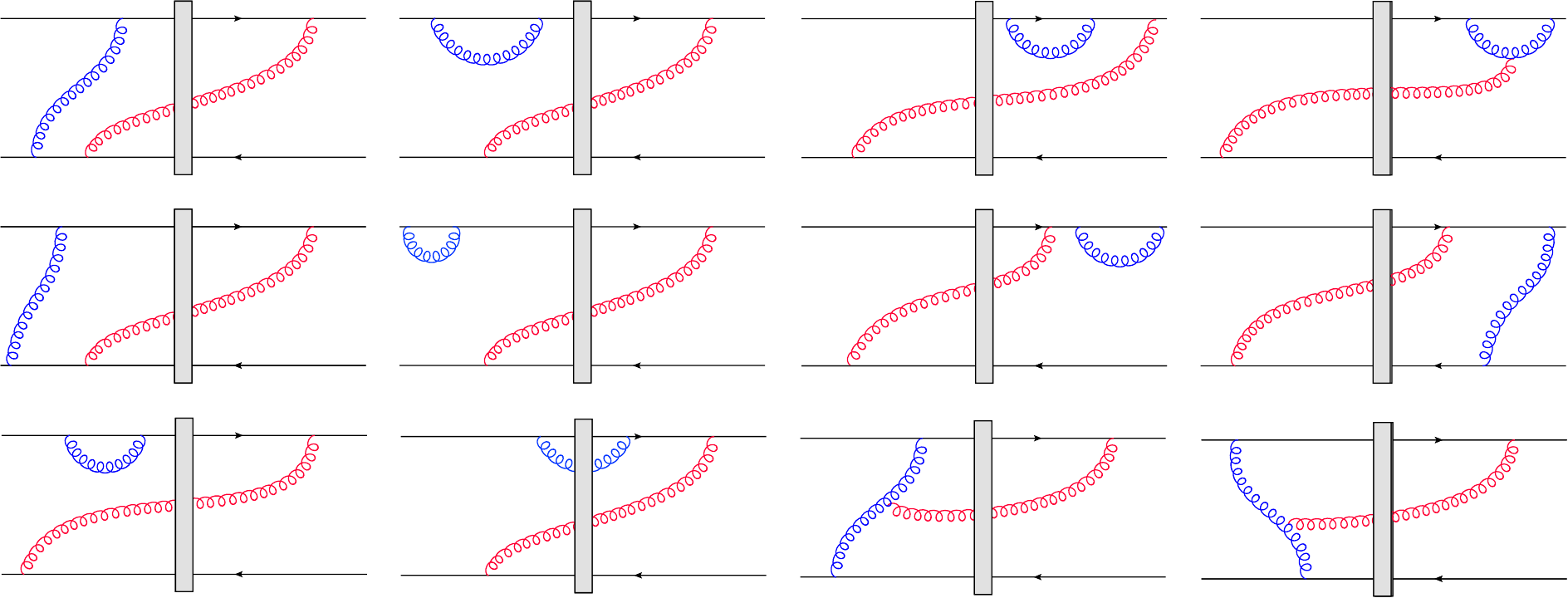}
\caption{Anti-time-ordered diagrams ($\tau_p<\tau_k$).}
\label{fig2}
\end{figure}
A quite non-trivial fact is that a number of anti-time ordered graphs (Fig. \ref{fig2}), involving factors $\frac{p^-}{p^-+k^-}\simeq \Theta (\tau_k-\tau_p)$, which would also be potentially enhanced by double transverse logs since
\begin{equation}
 \int_{r^2}^{\bar{z}^2}\frac{\rmd\bar{u}^2}{\bar{u}^2}\int^{q^+}_{k^+\frac{\bar{z}^2}{\bar{u}^2}}\frac{\rmd p^+}{p^+}=\int_{r^2}^{\bar{z}^2}\frac{\rmd\bar{u}^2}{\bar{u}^2}\ln\frac{\bar{z}^2}{\bar{u}^2}=\frac{\rho^2}{2},
\end{equation}
cancel completely among themselves\footnote{Although virtual contributions do not involve a natural time ordering and are not enhanced by double logs, it is convenient to artificially split them in time- and anti-time-ordered contributions. In this way they regulate UV and IR divergences for both time orderings independently.}. Thus we conclude that perturbative corrections enhanced by double logarithms $Y\rho$ or $\rho^2$ can be resummed to all orders by solving a modified DLA equation involving manifest time-ordering
\begin{equation}\label{eq:nonlocal}
 {\cal A}(q^+,r^2)={\cal A}(0,r^2)+\asbar\int_{r^2}^{1/Q_0^2}\frac{\rmd z^2}{z^2}\int_{q_0^+}^{q^+\frac{r^2}{z^2}}\frac{\rmd k^+}{k^+}{\cal A}(k^+,z^2).
\end{equation}

\section{The Resummed BK Evolution Kernel}\label{3}

Eq. \eqref{eq:nonlocal} is manifestly non-local in rapidity, being equivalent to $\partial_Y {\cal A}(Y,\rho)=\asbar\int_0^\rho\rmd\rho_1{\cal A}(Y-\rho+\rho_1,\rho).$
Nevertheless, direct iteration of this equation and some mathematical manipulations allows us to write \cite{ours}
\begin{equation}\label{eq:trick}
 \tilde{\cal A}(Y,\rho)\equiv\int_0^\rho\rmd\rho_1\tilde{f}(Y,\rho-\rho_1){\cal A}(0,\rho_1),\qquad \tilde{f}(Y,\rho)=\int_{\cal C}\frac{\rmd\gamma}{2\pi i}J(\gamma)\rme^{\left[\asbar\chi_{\rm DLA}(\gamma)Y+(1-\gamma)\rho\right]}.
\end{equation}
Here $\tilde{\cal A}(Y,\rho)$ is the analytical continuation of the physical amplitude ${\cal A}(Y,\rho)$ for $Y>\rho$, and
\begin{equation}
 \asbar\chi_{\rm DLA}(Y)=\frac{1}{2}\left[-(1-\gamma)+\sqrt{(1-\gamma)^2+4\asbar}\right],\quad J(\gamma)=1-\asbar\chi'_{\rm DLA}(\gamma).
\end{equation}
Eq. \eqref{eq:trick} implies that we can write the DLA-resummed evolution of the amplitude in terms of a  $Y$-independent kernel ${\cal K}_{\rm DLA}(\rho)$ (the inverse Mellin transform of the characteristic function $\chi_{\rm DLA}(\gamma)$):
\begin{equation}\label{kernel}
 \tilde{A}(Y,\rho)=\tilde{A}(0,\rho)+\asbar\int_0^Y\rmd Y_1\int_0^\rho\rmd \rho_1{\cal K}_{\rm DLA}(\rho-\rho_1){\tilde A}(Y_1,\rho_1), \quad{\cal K}_{\rm DLA}(\rho)=\frac{J_1(2\sqrt{\asbar\rho^2})}{\sqrt{\asbar\rho^2}}.
\end{equation}
Notice that the resummation also enters the initial condition, given by the limit of \eqref{eq:trick} at the analytically continued point $Y=0$, $\tilde{f}(0,\rho)=\delta(\rho)-\sqrt{\asbar}J_1(2\sqrt{\asbar\rho^2})$, from which analytical expressions for $\tilde{A}(0,\rho)$ can be obtained for initial conditions of the Golec-Biernat-Wüsthoff (${\cal A}(0,\rho)_{\rm GBW} \sim 1$) \cite{GBW} and McLerran-Venugopalan (${\cal A}(0,\rho)_{\rm MV}\sim \rho$) \cite{MV} type. After having performed the resummation at double-logarithmic accuracy, it is easy to promote Eq. \eqref{kernel} into a more complete equation matching NLO BK\footnote{Single-log (collinear and running-coupling) and finite terms appearing in NLO BK have not been included in Eq. \eqref{promotion}, although they can be simply added to the kernel. Prescriptions to resum single logs are discussed in \cite{fit}.} \cite{ours}:
\begin{equation}\label{promotion}
 \partial_Y T_{\bx\by}=\frac{\asbar}{2\pi}\int \rmd^2\bz\,{\cal M}_{\bx\by\bz}{\cal K}_{\rm DLA}\left(\sqrt{\ln\tfrac{(\bx-\bz)^2}{(\bx-\by)^2}\ln\tfrac{(\by-\bz)^2}{(\bx-\by)^2}}\right)[T_{\bx\bz}+T_{\bz\by}-T_{\bx\by}-T_{\bx\bz}T_{\bz\by}],\quad Y>\rho.
\end{equation}

\section{Some Numerical Results}\label{4}

\begin{figure}[t]
\centering
\includegraphics[scale=0.7]{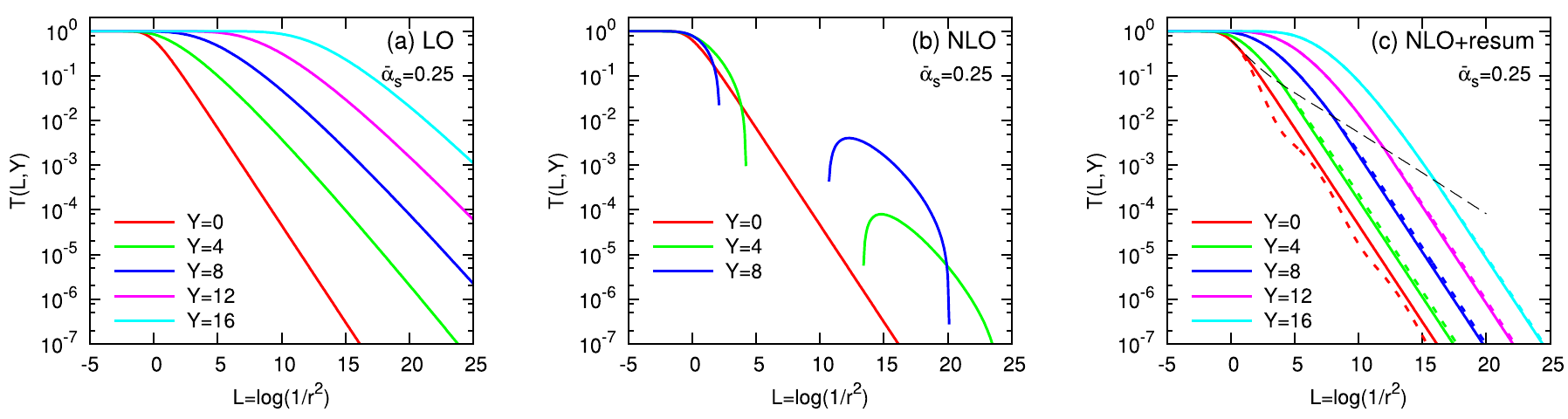}
\caption{Numerical solutions to BK equation for the dipole amplitude $(\asbar=0.25, {\cal A}(0,\rho)=1)$ at strict LO (left), including the NLO double log (center), and the resummed tower of double logs to all orders (right).}
\label{fig3}
\end{figure}
\begin{figure}[t]
\centering
\includegraphics[scale=0.6]{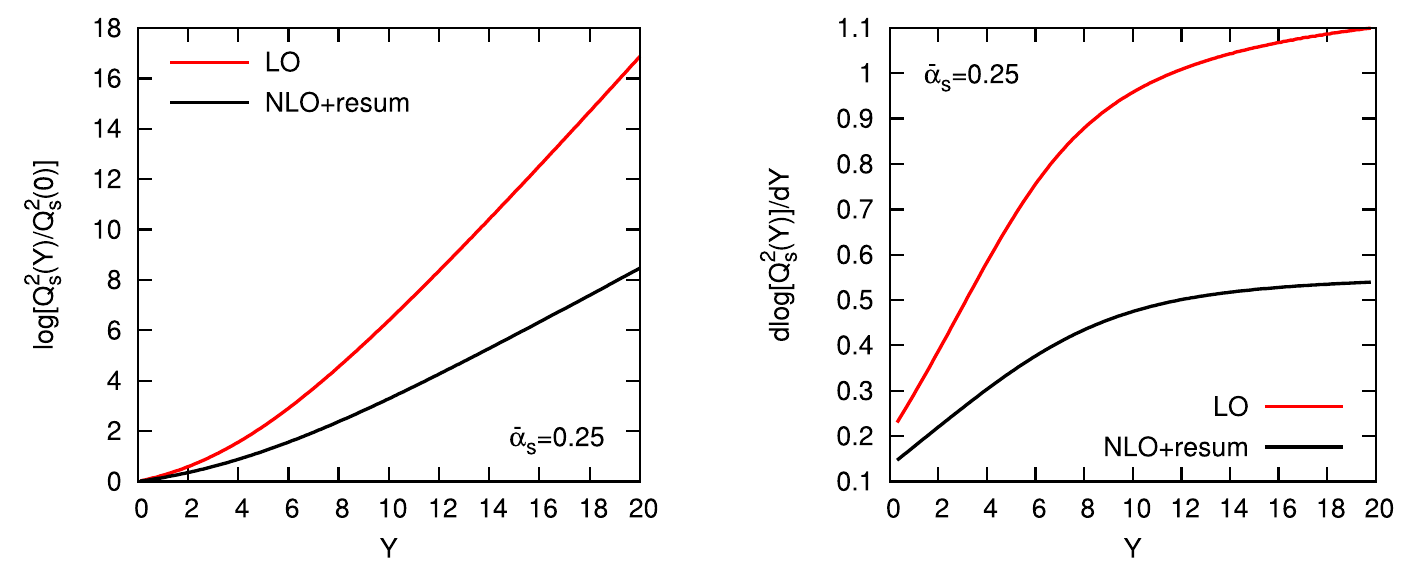}
\caption{Rapidity dependence of the target saturation momentum $Q_s^2(Y)$, with $\asbar=0.25$.}
\label{fig4}
\end{figure}

The numerical solution to Eq. \eqref{promotion} is presented in Fig. \ref{fig3}. We clearly see that while introducing the double log term from the NLO BK kernel renders the evolution unstable and physically meaningless, the evolution is perfectly smooth after resummation and certainly slower than that observed at LO. To characterize more precisely this reduction, we computed numerically the target saturation momentum $Q_s^2(Y)$ (Fig. \ref{fig4}). The growth of the saturation scale with $Y$ is considerably reduced by the resummation: for sufficiently large rapidities, the saturation exponent $\lambda_s\equiv\rmd(\ln Q_s^2(Y)/Q_0^2)/\rmd Y$ is roughly reduced by a factor of 2 for the resummed kernel as compared to LO one. This has important phenomenological consequences that have begun to be explored in \cite{fit}, where also the relevant issue of single logarithmic (running coupling and collinear) corrections is addressed.

\end{document}